\documentstyle[12pt,aasms4]{article}

\lefthead{A. R. Walker et al. }
\righthead{The Distance of the Large Magellanic Cloud Cluster NGC 1866}

\newcommand{\lsim}{\ \raise -2.truept\hbox{\rlap{\hbox{$\sim$}}\raise5.truept\hb
ox{$<$}\ }} 
\newcommand{\gsim}{\ \raise -2.truept\hbox{\rlap{\hbox{$\sim$}}\raise5.truept\hb
ox{$>$}\ }}

\begin{document}

\title{The Distance of the Large Magellanic Cloud Cluster NGC 1866}

\author{A. R. Walker \altaffilmark{1}, G. Raimondo \altaffilmark{2} \altaffilmark{5}, 
E. Di Carlo \altaffilmark{2},  E. Brocato \altaffilmark{2},
V. Castellani \altaffilmark{3},
V. Hill \altaffilmark{4}}

\altaffiltext{1}{Cerro Tololo Inter-American Observatory, National Optical
Astronomy Observatory, 
Casilla 603, La Serena, Chile; awalker@noao.edu.  NOAO is operated by AURA 
Inc., under cooperative agreement with the National Science Foundation.}

\altaffiltext{2}{Osservatorio Astronomico di Collurania, Via M. Maggini, I-64100
Teramo, 
Italy; brocato, dicarlo, raimondo@astrte.te.astro.it}

\altaffiltext{3}{Dipartimento di Fisica, Universit\`a di Pisa, P.za Torricelli 2,
I-56100 Pisa, Italy; vittorio@astr18pi.difi.unipi.it}

\altaffiltext{4}{European Southern Observatory; vhill@eso.org}
\altaffiltext{5}{Astronomia-Dipartimento di  Fisica, Universit\`a La Sapienza,
P. le Aldo Moro 2, I-00185 Roma, Italy}

\begin{abstract} 

Hubble Space Telescope V, I photometry of stars in the Large
Magellanic Cloud cluster NGC 1866 shows a well defined cluster main
sequence down to V=25 mag, with little contamination from field or
foreground stars.  We use the main sequence fitting procedure to link
the distance of NGC 1866 to the Hipparcos determination of the
distance for the Hyades MS stars, making use of evolutionary
prescriptions to allow for differences in the chemical composition.
On this basis we find a true distance modulus for NGC 1866 of $18.35
\pm 0.05$ mag.  If the cluster is assumed to lie in the LMC plane then
the LMC modulus is 0.02 mag less.

\end{abstract}

 
\section{Introduction}  

The distance to the Large Magellanic Cloud (LMC) is a critical step in
the establishment of the distance scale, since it allows us to compare
and thus cross-calibrate a variety of methods, and on the basis of
this evaluation the identification of reliable distance indicators
should follow.  Only then can extension to the more distant Universe
be confidently undertaken. However, the distance modulus (DM) of the
LMC is still controversial since estimations from various indicators
cover the range 18.2$-$18.7 mag (Walker 1999), and there is no
definitive measurement available that could settle this dispute.  We
attempt here to improve this situation by providing an accurate DM for
the Cepheid-rich LMC cluster NGC 1866 via the technique of
main-sequence (MS) fitting.

NGC 1866 is a populous young cluster sited some $4^{\circ}$ north of
center of the LMC.  From the time of the pioneering work by Arp \&
Thackeray (1967) it has served as a laboratory for stellar evolution
studies of intermediate mass $(\sim 5M_{\odot})$ stars, as the cluster
is sufficiently rich that significant numbers of stars appear in rare
stages of evolution, these include at least 20 Cepheids (Welch \&
Stetson 1993 and references therein).  Although several efforts have
been made to secure high-quality photometry for NGC 1866, most
recently by Testa et al.  (1999), ground-based efforts are hampered by
crowding, and by contamination from LMC field stars.  Consequently, we
have observed NGC 1866 with Hubble Space Telescope (HST) WFPC2,
allowing accurate photometry several magnitudes down the main
sequence, together with greatly reduced sensitivity to crowding and
contamination.

\begin{figure*}[t]
\centerline{
\epsfxsize=9cm\epsfxsize=9cm\epsfbox{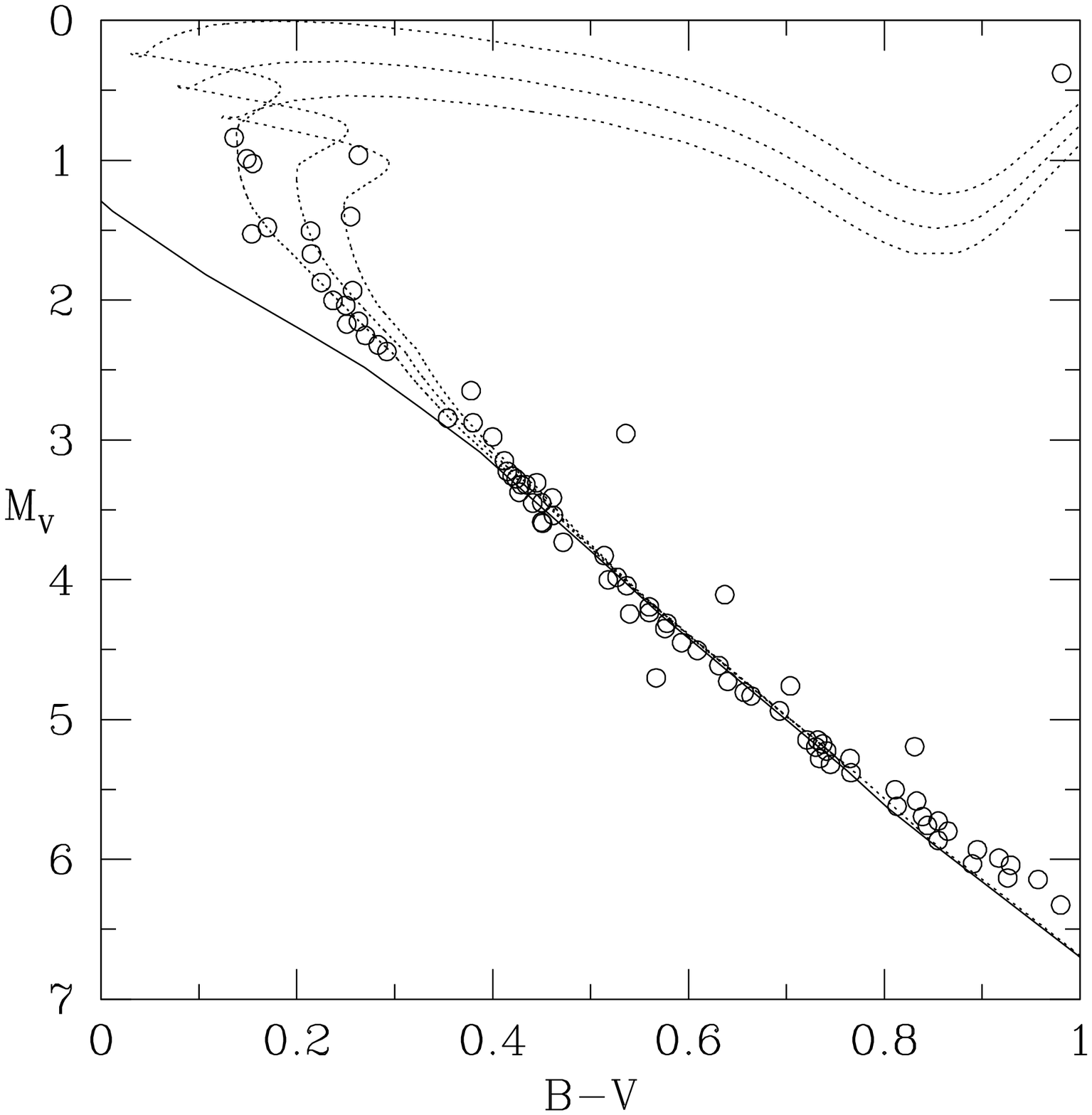}
\hspace{-.25cm}\epsfxsize=9cm\epsfbox{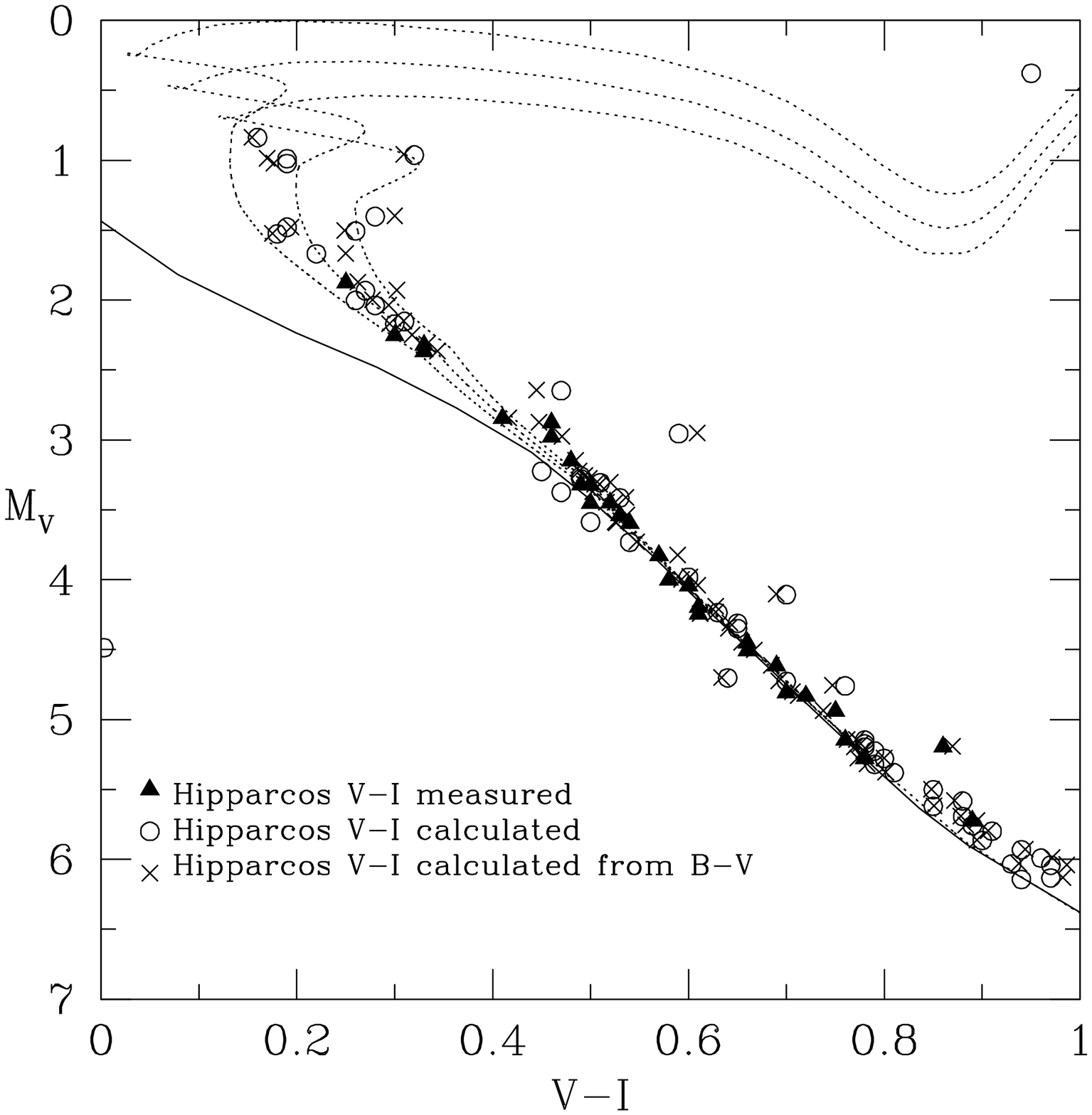}}
\figcaption{The M$_{V}$,B$-$V ({\it Panel a}) and M$_{V}$,V$-$I 
({\it Panel b}) diagrams of the subsample of 
the stars of the Hyades as provided with corrected parallaxes by Madsen S., 
Dravins D., Lindegren L. (2001). The solid lines represent the
 zero age theoretical models computed by adopting [Fe/H]=+0.13. 
Three isochrones aged 500, 600 and 700 Myr are also plotted (dotted lines).
\label{fig1}
}
\end{figure*}

A detailed presentation of the observations, and comparisons with
evolutionary theory will be made elsewhere (Brocato et al., in
preparation).  In section 2 we describe the observations and the
photometric calibration, in section 3 we test the correctness of our
method of tying the NGC 1866 MS to Hipparcos parallaxes, in section 4
we fit to the NGC 1866 MS, and in section 5 summarize the results of
the analysis.

\section{Observations and Photometric Calibration}

The WFPC2 data set consists of two sets of pointings, one with NGC
1866 centered on WF3, and the other on PC1, through V (F555W) and I
(F814W) filters.  Three sets of different exposures times were taken,
with multiple exposures for each.  Photometry was performed using the
program CCDCAP (Mighell et al. 1996), followed by conversion to the
standard photometric system (Johnson V, Cousins I) via equations and
zeropoints listed by Holtzman et al. (1995).  Corrections for
geometric distortion were also applied, together with CTE corrections
according to prescriptions by Whitmore, Heyer, \& Casertano (1999),
with the exception that no long-short correction was applied, as our
tests on this and other datasets using CCDCAP have found such a
correction to be unnecessary (see also Dolphin 2000).  The several
datasets were matched and combined, and brought to an internally
consistent system.

As the WFPC2 photometric zeropoints are uncertain at the $\sim 0.02$
magnitude level (Holtzman et al. 1995, Dolphin 2000), we compared our
photometry for the merged dataset to the ground-based CMD by Walker
(1995), which is referenced to a sequence of local standards in the
vicinity of NGC 1866, that are in turn tied in to the standard
Johnson-Cousins system to $\pm0.01$ mag in both V magnitude and B$-$V
and V$-$I colors.  We based our comparison on all available overlap
stars ($\sim 250$), to find the differences ground-HST, $\Delta$V$ =
0.007 \pm 0.09$ (s.d.), $\Delta$(V$-$I)$ = -0.07 \pm 0.06$ (s.d.);
trimming the sample to within $\pm 0.1$ mag of the mean changed both
differences to $\Delta$ = 0.01 mag and reduced the s.d. by a factor of
two.  No systematic differences as a function of color or magnitude
were found, and given the small size of the corrections, we did not
adjust the HST photometry.

\section{From Hyades by HIPPARCOS to NGC 1866 by HST} 

We wish to relate the NGC 1866 MS to the system of Hipparcos
parallaxes (ESA 1997) with the minimum number of steps and
assumptions.  We chose to use the Hyades as our fundamental fiducial,
and calculated the absolute magnitudes individually using the new
kinematically improved parallaxes, where the error in the Hipparcos
catalog has been diminished by combining its data with a kinematic
modeling of the cluster dynamics (Madsen, Lindegren \& Dravins 2000;
Madsen, Dravins \& Lindegren 2001).  The binary systems identified by
Perryman et al. (1998) are excluded, the sample totals 111 stars and
has a well-determined mean DM of $3.33 \pm 0.01$ mag.  The controversy
over whether or not correlated errors systematically affect Hipparcos
distances to nearby open clusters is irrelevant for the Hyades due to
compensating effects (Pinsonneault, Terndrup, \& Yuan 2000).

For our purposes, a complication is that many stars in the Hipparcos
catalog, including most of the Hyades, do not have V$-$I colors
actually measured on the Cousins system, instead a variety of
transformations are applied, depending on available photometry, to
produce a quasi-Cousins V$-$I.  Of our 111 Hyades stars, 83 are MS
members with V$-$I $< 1.0$.  For these 83 stars, 29 have photometry
actually measured on the Johnson-Cousins system, the remainder have
V$-$I colors calculated as described in the Hipparcos catalog (ESA
1997).  The extreme tightness of the Hyades MS is evidence that the
transformation procedure works well, as can be seen by comparing the
$M_V$, B$-$V and $M_V$, V$-$I CMDs, plotted in Figure 1.  On Figure 1b
we plot the transformed B$-$V colors according to the precepts of
Cousins (1978) (crosses), to demonstrate the concept developed in
greater detail in the Hipparcos analysis.  We differentiate, using
different symbols, between the stars with measured V$-$I Cousins and
those with transformed colors.  This comparison shows no indication of
significant systematic differences for the non-evolved stars over
our color range of interest.

The critical step of comparing the MS of the Hyades with that for NGC
1866 requires a reliable comparison method, and accurate metallicties
for both clusters.  Recent evaluations of the Hyades metallicity are
all very consistent, with [Fe/H] = 0.12 $\pm$ 0.03 (Cayrel et
al. 1985), [Fe/H] = 0.13 $\pm$ 0.02 (Boesgaard \& Friel 1990) [Fe/H] =
0.14 $\pm$ 0.05 (Perryman et al. 1998).  We adopted [Fe/H] = 0.13 and
combined the helium abundance with the metallicity according to the
relation $\Delta Y$/$\Delta Z$$\sim$2 with Y=0.27 and Z=0.02 for the
Sun. Thus we will assume Y=0.282 and Z=0.026 for the Hyades.  We note
that Perryman et al. (1998) and Castellani et al. (2001) used slightly
lower metallicity (Z=0.024) on the basis of a different assumption on
the solar ratio (Z/H). However such a difference in metallicity
corresponds to a negligible shift in the location of the zero age
stellar models in the color-magnitude diagram due to the corresponding
decrease of the helium content.

We accommodate the metallicity difference between the Hyades and NGC
1866 by computing a set of stellar models for the mass range $0.7$ to
9M$_{\odot}$, chemical composition (Y=0.282 and Z=0.026) and a mixing
length parameter $\alpha $=2.0 using the evolutionary code FRANEC.
The present version of the code uses the most recent physical inputs,
in particular the OPAL equation of state and opacity (Cassisi et al.
1998). Neither diffusion nor $\alpha $-enhancement are adopted.
Atmosphere models are from Castelli (1997) compilation
(http://cfaku5.harvard.edu/grids.html) computed without any
overshooting, see for a discussion Castelli, Gratton, \& Kurucz 1997.

Taylor (1980) found a negligible value of E(B$-$V) =0.003$\pm$0.002
for the Hyades reddening, so we adopt zero reddening correction.

The computed ZAMS is plotted in figure 1, to show that there is
excellent agreement over most of the range of the non-evolved stars
(M$_{V}>$3 mag).  We also plot a sample of isochrones calculated 
at three different ages (for a discussion of the Hyades age see
Castellani et al. 2001 and references therein).  The models are 
slightly bluer than the MS only for the very reddest stars.  We
conclude that our models correctly describe the Hyades MS, and
in particular the ZAMS model is an excellent fit for
$0.5 < $V$-$I$ < 0.8$.  On the basis of this result, we
can procede with confidence to fit our ZAMS models to the NGC 
1866 CMD.  We note that shifting the Hyades MS to the  
NGC 1866 metallicity $-$ i.e. using theory in a differential way $-$,
is exactly the same as directly comparing a new theoretical ZAMS,
computed with the NGC 1866 chemical composition, to the NGC 1866 CMD.
Additionally, since the method makes use of zero age main sequence
models, the differences between evolutionary tracks provided by
different groups are only a minor source of indetermination.

\section{The NGC 1866 main sequence fitting}

The NGC  1866 MS plotted  in Figure~2 shows  a very clearly  defined and
well-populated sequence of single  stars, with two significant changes
in slope  near V=21 and V=22  which will strongly  constrain the model
fit; in  this respect the V,  V$-$I CMD has a  distinct advantage over
the V,  B$-$V CMD.   The region  V$-$I $> 0.8$,  where our  Hyades fit
deviates slightly,  has little power in  the fit.  There  is clearly a
significant population  of binaries, as suggested  previously by Testa
et  al. (1999).   Contamination by  the older  field  star population,
visible as a RGB and RG  clump, with turnoff at V$\sim23$, is minimal.
We calculated the expected younger field star contamination to the NGC
1866 MS by  scaling field star photometry from  Walker (1995).  In the
range V=19-20, V$-$I= $-$0.05-0.3, we expect 28 field stars on our HST
frames.  Since  we find 1074 stars in  this range on our  HST CMD, the
field star  contamination in the vicinity  of the NGC 1866  MS is very
small, and will have negligible effect on our fits.


Recent evaluations of metallicity of NGC 1866 via Stromgren photometry
of a few stars found the value [Fe/H] = $-$0.43$\pm$0.18 (Hilker et
al. 1995) and from the integrated spectrum Oliva \& Origlia (1998)
obtained [Fe/H] = $-$0.55$\pm$0.4.  Using the ESO VLT with the high
dispersion spectrograph UVES, Hill et al. (2000) have measured
abundances for three NGC 1866 RGB stars, finding [Fe/H] =
$-$0.50$\pm$0.1 and [$\alpha $/Fe] = +0.1$\pm$0.1 for O, Mg, Ti and Ca
elements.  The internal scatter for the three [Fe/H] values is only
0.05 dex.

With the same version of the stellar evolutionary code and
prescriptions described above, we computed a new set of stellar models
for two metallicities Z=0.007, which corresponds to [Fe/H] = $-0.50$,
and a higher value Z=0.01 (around [Fe/H] = $-0.30$).  The helium
abundance is calculated by the above relation $\Delta Y$/$\Delta
Z$$\sim$2 as recently confirmed for the Small Magellanic Cloud by
Peimbert, Peimbert, \& Ruiz (2000), so respectively Y=0.24 and 0.25.
The computed ZAMS models for these Z values are very consistent 
with Castellani, Degl'Innocenti \& Marconi (1999)
and with models computed for solar scaled abundances 
recently published by Salasnich et al. (2000) at the closest 
metallicity (Z=0.008 Y=0.25).

A set of models computed for an higher (Y=0.27) and lower (Y=0.23)
original helium abundance at the metallicity Z=0.007 disclose that the
ZAMS becomes respectively fainter and brighter by about 0.05 mag in
M$_{V}$ in the relevant V$-$I color range.

To perform an accurate fit we derived a fiducial line for the portion
of the MS ranging from about V=25 mag up to 20 mag.  The last point is
estimated by superimposing a sample of suitable isochrones with
different ages looking for the magnitude level where the isochrones
turn away from the zero age main sequence.  In this way we are
confident to avoid any contamination by stars evolved off the ZAMS.
It is also important to note that the CMD is sufficiently deep and 
accurate such that the fit over the precise range of V$-$I colors where
the Hyades ZAMS matches so well is identical (but with larger photometric
error) to the fit using all the non-evolved NGC 1866 stars.

The points of the fiducial line have been derived with a running mean
technique by taking the maximum value in the V$-$I histogram       
within each bin of magnitude.  Then, we apply the MS fitting method
comparing the theoretical ZAMSs with the observed fiducial line. By
minimizing the $\chi^{2}$ we obtain (m-M)$_{V}$ = 18.50$\pm$0.05 and
E(V$-$I) = 0.08$\pm$0.01 with Z=0.007 and (m=M)$_{V}$ = 18.53$\pm$0.05
and E(V$-$I) = 0.075$\pm$0.01 for Z=0.01.  This procedure allows to
derive separately both reddening and distance.  The errors refer to
the uncertainties due to the method adopted to built the fiducial line
(i.e. the bin size in magnitude and color, the amplitude of the
running mean and the range in V considered for the fit).

By assuming $R_V = A_V/E_{B-V} = 3.1$, Bessell \& Brett (1988) found
the relation E(V$-$I) = 1.25E(B$-$V) thus the reddening values
E(V$-$I)=0.08 and E(V$-$I)=0.075 imply respectively E(B$-$V)=0.064 and
E(B$-$V)=0.06, in agreement with the evaluations in the vicinity of
NGC 1866 derived from UBV photometry, E(B$-$V) = 0.060 $\pm$ 0.005
(Arp 1967, van den Bergh \& Hagen 1968; Walker 1974). We note that the
reanalysis of the ultraviolet extinction in LMC by Misselt et
al. (1999) suggests on average $R_{V}$=2.6; if this value is used then
distances increase by only 0.03 mag.  Taking into account the
uncertainty due to the chemical composition we suggest that the main
sequence fitting method applied to the cluster NGC 1866 gives an
absolute DM of (m$-$M)$_{0}$ = 18.35$\pm$0.05 (1$\sigma$).  If NGC
1866 is assumed to lie in the plane of the LMC then the correction to
the LMC center is $-0.02$ mag, thus we derive a DM for the LMC of
(m$-$M)$_{0}^{LMC}$ = 18.33$\pm$0.05 mag.

\section{Final Remarks}

In this work we have determined a distance to the LMC based on the
well-defined Hipparcos distance to the Hyades, using theoretical
models to account for the metallicity difference.  With high-quality
photometry and accurate abundances, the method appears robust.  It
would be very valuable to test the technique on equivalent data for
other young LMC clusters, over a range of [Fe/H].

The result here is consistent with the infrared surface-brightness DM
of $18.42 \pm 0.10$ (Gieren et al. 2000) for a single NGC 1866 Cepheid
(HV 12198); further results of this type for more NGC 1866 Cepheids
are expected soon (W. Gieren, private communication) which should
allow a more critical comparison.
 
Since NGC 1866 contains a large number of Cepheids, the accurate study
of their properties provides a unique opportunity to link stellar
evolution theory and pulsational models and to evaluate both the
distance of the LMC and the degree of confidence in the Cepheid PL and
PLC relations, which are fundamental steps in the building of the
cosmological distance scale.

\section{Acknowledgements}

We warmly thank Dainis Dravins for providing us the corrected
Hipparcos parallaxes before publication, and Ken Mighell for supplying
aperture corrections and other help with CCDCAP. One of us (G.R.)
acknowledges support during a visit to CTIO, and from a grant of the
Consorzio Nazionale di Astronomia e Astrofisica (CNAA).  This work is
supported by the Italian Ministry of University, Scientific Research
and Technology (MURST): Cofin2000-Project: {\sl 
Stellar Observables of Cosmological Relevance},
and by HST Grant GO-08151.01-97A.


\begin{thebibliography}{}
\bibitem[]{} Arp, H., 1967, 149, 73
\bibitem[]{} Arp, H., Thackeray, A. D., 1967, ApJ 149, 7 
\bibitem[]{} Bessell, M. S., Brett, J. M. 1988, PASP, 100, 1134
\bibitem[]{} Boesgaard, A. M., \& Friel, E. D., 1990, ApJ, 351, 467
\bibitem[]{} Cassisi, S., Castellani, V., Degl'Innocenti, S., Weiss, A. 1998,
A\&AS, 129, 267
\bibitem[]{} Castellani, V., Degl'Innocenti, S., Marconi, M., 1999, A\&A 349, 834
\bibitem[]{} Castellani, V., Degl'Innocenti, S., Prada Moroni, P. G.,  2001, MNRAS 320, 66
\bibitem[]{} Castelli, F., Gratton R., Kurucz L. R., 1997, A\&A 318, 841
\bibitem[]{} Cayrel, R., Cayrel de Strobel, G., Campbell, B., 1985, A\&A, 146, 294
\bibitem[]{} Cousins, A. W. J., 1978, MNASSA, 37, 62
\bibitem[]{} Dolphin, A. E., 2000, PASP, 112, 1397
\bibitem[]{} ESA, 1997, The Hipparcos and Tycho Catalogues, ESA SP-1200
\bibitem[]{} Gieren, W. P., Storm, J., Fouqu\'e, P., Mennickent, R. E., G\'omez, M., 2000, ApJ 533, 107
\bibitem[]{} Hilker, M., Richtler, T., Gieren, W., 1995, A\&A 294, 648
\bibitem[]{} Hill, V., Francois, P., Spite, M., Primas, F., Spite,F., 2000, A\&A, 364, L19
\bibitem[]{} Holtzman, J.A., et al., 1995, PASP, 107, 1065
\bibitem[]{} Madsen, S., Dravins, D., Lindegren, L., 2001 A\&A, to be submitted
\bibitem[]{} Madsen, S., Lindegren, L., Dravins, D., 2000, ASPC 198, 137 
\bibitem[]{} Mighell, K. J., Rich, R. M., Shara, M., Fall, M., 1996, AJ, 111, 2314
\bibitem[]{} Misselt, K. A., Clayton, G. C., Gordon, K. D., 1999, ApJ 515, 128
\bibitem[]{} Oliva, E., Origlia, L., 1998, A\&A 332, 46
\bibitem[]{} Peimbert, M., Peimbert, A., Ruiz, M. T., 2000, ApJ 541, 688
\bibitem[]{} Perryman, M. A. C., et al., 1998, A\&A 331, 81
\bibitem[]{} Pinsonneault, M. H., Terndrup, D.M., \& Yuan, Y., 2000, in "Stellar Clusters
and Associations: Convection, Rotation, and Dynamos", eds. R. Pallavicini, G. Mirela,
and S. Sciortino, ASP Conf. Ser. 198, 95
\bibitem[]{} Salasnich, B., Girardi, L., Weiss, A., Chiosi, C., 2000, A\&A 361, 1023
\bibitem[]{} Taylor , 1980, AJ 85, 242
\bibitem[]{} Testa, V., Ferraro, F. R., Chieffi, A., Straniero, O., Limongi, M., Fusi Pecci, F., 1999, AJ 118, 2839
\bibitem[]{} Van den Bergh, S., \& Hagen, G. L., 1968, AJ, 73, 569
\bibitem[]{} Walker, A. R., 1995, AJ 110, 638
\bibitem[]{} Walker, A. R., 1999, in "Post Hipparcos Cosmic Candles", ed. A. Heck
and F. Caputo, Kluwer Academic Publishing, p125
\bibitem[]{} Walker, M. F., 1974, MNRAS, 169, 199
\bibitem[]{} Welch, D. L., Stetson, P. B., 1993, AJ 105, 1813
\bibitem[]{} Whitmore, B., Heyer, I., Casertano, S., 1999, PASP, 111, 1559
\end{thebibliography}
\end{document}